\documentclass[aps, prl, twocolumn, amsmath, amssymb, showpacs, superscriptaddress
]{revtex4-1}

\usepackage[utf8]{inputenc}
\usepackage{graphicx}
\usepackage{units}
\usepackage{color}
\usepackage[pdftex, colorlinks=true, linkcolor=myblue, citecolor=myblue,
urlcolor=myblue]{hyperref}
\usepackage{times}
\usepackage{bbold}

\definecolor{myblue}{rgb}{0,0,1}

\begin{document}

\title{Dirac-like plasmons in honeycomb lattices of metallic nanoparticles}

\author{Guillaume Weick}
\affiliation{Institut de Physique et Chimie des Mat\'eriaux de Strasbourg,
Universit\'e de Strasbourg, CNRS UMR 7504, 23 rue du Loess, BP 43, F-67034
Strasbourg Cedex 2, France}

\author{Claire Woollacott}
\affiliation{Centre for Graphene Science, Department of Physics and Astronomy, University of
Exeter, Stocker Rd.\ EX4 4QL Exeter, UK}

\author{William L. Barnes}
\affiliation{Centre for Graphene Science, Department of Physics and Astronomy, University of
Exeter, Stocker Rd.\ EX4 4QL Exeter, UK}

\author{Ortwin Hess}
\affiliation{The Blackett Laboratory, Department of Physics, Imperial College London,
South Kensington Campus, London SW7 2AZ, UK}

\author{Eros Mariani}
\affiliation{Centre for Graphene Science, Department of Physics and Astronomy, University of
Exeter, Stocker Rd.\ EX4 4QL Exeter, UK}


\begin{abstract}
We consider a two-dimensional honeycomb lattice of metallic nanoparticles, each
supporting a localized surface plasmon, and study the quantum properties of the
collective plasmons resulting from the near field dipolar interaction between
the nanoparticles. We analytically investigate the dispersion, the effective
Hamiltonian and the eigenstates of the collective plasmons for an arbitrary
orientation of the individual dipole moments.
When the polarization points close to the normal to the plane the spectrum
presents Dirac cones, similar to those present in the electronic band structure
of graphene. 
We derive the effective Dirac Hamiltonian for the collective plasmons and show 
that the corresponding spinor eigenstates 
represent Dirac-like massless bosonic excitations
that present similar effects to electrons in graphene, such as a
non-trivial Berry phase and the absence of backscattering off smooth inhomogeneities.
We further discuss how one can manipulate the Dirac points in the Brillouin zone
and open a gap in the collective plasmon dispersion by modifying the
polarization of the localized surface plasmons, paving the way for a
fully tunable plasmonic analogue of graphene.
\end{abstract}


\maketitle

Light has been the source of inspiration for scientific thinking for
millennia. Ancient Assyrians developed the first lenses in order to bend the
trajectory of light and control its propagation. In contrast to the macroscopic
scale, the use of light to observe microscopic structures poses difficulties
due to the diffraction limit \cite{abbe}. In an attempt to overcome this limit
and observe subwavelength structures, plasmonic nanostructures have been created
\cite{barne03_Nature, maier}, like isolated metallic nanoparticles 
\cite{klar98_PRL}. The evanescent field at the surface of the nanoparticle, associated to
the localized surface plasmon (LSP) resonance \cite{mie08_AP}, produces strong
optical field enhancement in the subwavelength region, allowing one to overcome the
diffraction limit and achieve resolution at the molecular level
\cite{kneip97_PRL}.

While the field of plasmonics mostly focuses on single or few structures, the
creation of ordered arrays of nanoparticles constitutes a bridge to the realm of
metamaterials. 
Plasmonic metamaterials exhibit unique properties beyond 
traditional optics, like negative refractive index \cite{vesel68_SPU}, perfect
lensing \cite{pendr00_PRL}, the exciting perspective of
electromagnetic invisibility cloaking \cite{leonh06_Science}, and ``trapped rainbow"
slow light exploiting the inherent broadband nature of plasmonics \cite{tsakm07_Nature}. Indeed, in plasmonic
metamaterials the interaction between LSPs on individual nanoparticles generates
extended plasmonic modes involving all LSPs at
once \cite{quint98_OL, brong00_PRB}. Understanding the nature and properties of these
plasmonic modes [referred to as ``collective plasmons" (CPs) in what follows]
is of crucial importance as they are the channel guiding electromagnetic
radiation with strong lateral confinement over macroscopic distances.

CPs in periodic arrays of metallic nanoparticles are an active area of
research in plasmonics because the interaction of the LSP resonances can lead to dramatic
changes in the overall optical response of such structures. For example, it was both
predicted \cite{carro86_JOSAB} and observed \cite{krave08_PRL}
that the plasmonic response of a periodic array of nanoparticles
could be significantly narrowed with respect to the single particle response.
Further work has shown that these coupled resonances are
relevant to applications in light emission
\cite{rodri12_APL}. The extended nature of the collective resonances
means that there is scope for harvesting emission from sources spread over large volumes
\cite{rodri11_PRX, atwat10_NatureMat}.

The dispersion of CPs and their physical nature crucially depend on the lattice
structure of the metamaterial and on the microscopic interaction between LSPs.
A lattice which recently generated remarkable interest in the
condensed matter community is the honeycomb structure exhibited by graphene, a
two-dimensional (2D) monolayer of carbon atoms \cite{novos04_Science}. In the case of
graphene, the hopping of electrons between neighboring atoms gives rise to a
rich band structure characterized by the presence of fermionic massless Dirac
quasiparticles close to zero energy \cite{walla47_PR}. The chirality associated with
pseudo-relativistic Dirac fermions results in several of the remarkable
properties of graphene, such as a nontrivial Berry phase accumulated in parallel
transport \cite{novos05_Nature} and the suppression of electronic
backscattering from smooth scatterers \cite{cheia06_PRB}.
Undoubtedly, it would be exciting
to harvest the remarkable physical properties of electrons in graphene in suitably
designed plasmonic metamaterials 
by analyzing the Hamiltonian and the
consequent nature of CP eigenmodes in 2D honeycomb lattices of metallic
nanoparticles. This is the purpose of the present theoretical paper.

We analytically show how the problem of interacting LSPs in the honeycomb
structure can be mapped to the kinetic problem of electrons hopping in graphene,
yielding massless Dirac-like bosonic CPs in the vicinity of two Dirac points in the
Brillouin zone. The conical dispersion of classical
plasmons in a honeycomb lattice
of nanoparticles has been discussed numerically
in the past for out-of-plane or purely
in-plane polarization \cite{han09_PRL}.
In quite different physical systems (e.g., photonic crystals
\cite{halda08_PRL}, acoustic waves
in periodic hole arrays \cite{torre12_PRL}, cold atoms \cite{tarru12_Nature}), 
conical dispersions were also found in
``artificial graphene" due to the honeycomb symmetry.
However, the existence of a conical dispersion is not sufficient \cite{mei12_PRB} to prove
the physical analogy between quantum CPs in
honeycomb plasmonic lattices and electrons in graphene.
In order to achieve that, here we unveil the \textit{full Dirac
Hamiltonian} of
quantum CPs as well as the pseudospin structure of the CP eigenmodes for dipolar LSPs with 
\textit{arbitrary} orientation.
The existence of Dirac points is
robust for a small in-plane component of the polarization, where the system maps to strained graphene
\cite{suzuu02_PRB}, while
band gaps can emerge for increasing in-plane
polarization. At energies away from the Dirac point, van Hove
singularities
emerge in the CP density of states (DOS),
associated with Lifshitz
transitions in the topology of equipotential lines \cite{lifsh60_JETP}. Our analysis highlights the
physical nature of CP eigenmodes as well as the tunability of their
band structure and of the corresponding DOS with the polarization of light, which
can be crucial for enhancing the coupling of light with the plasmonic
metamaterial at different wavelengths.


\begin{figure}[tb]
\includegraphics[width=\columnwidth]{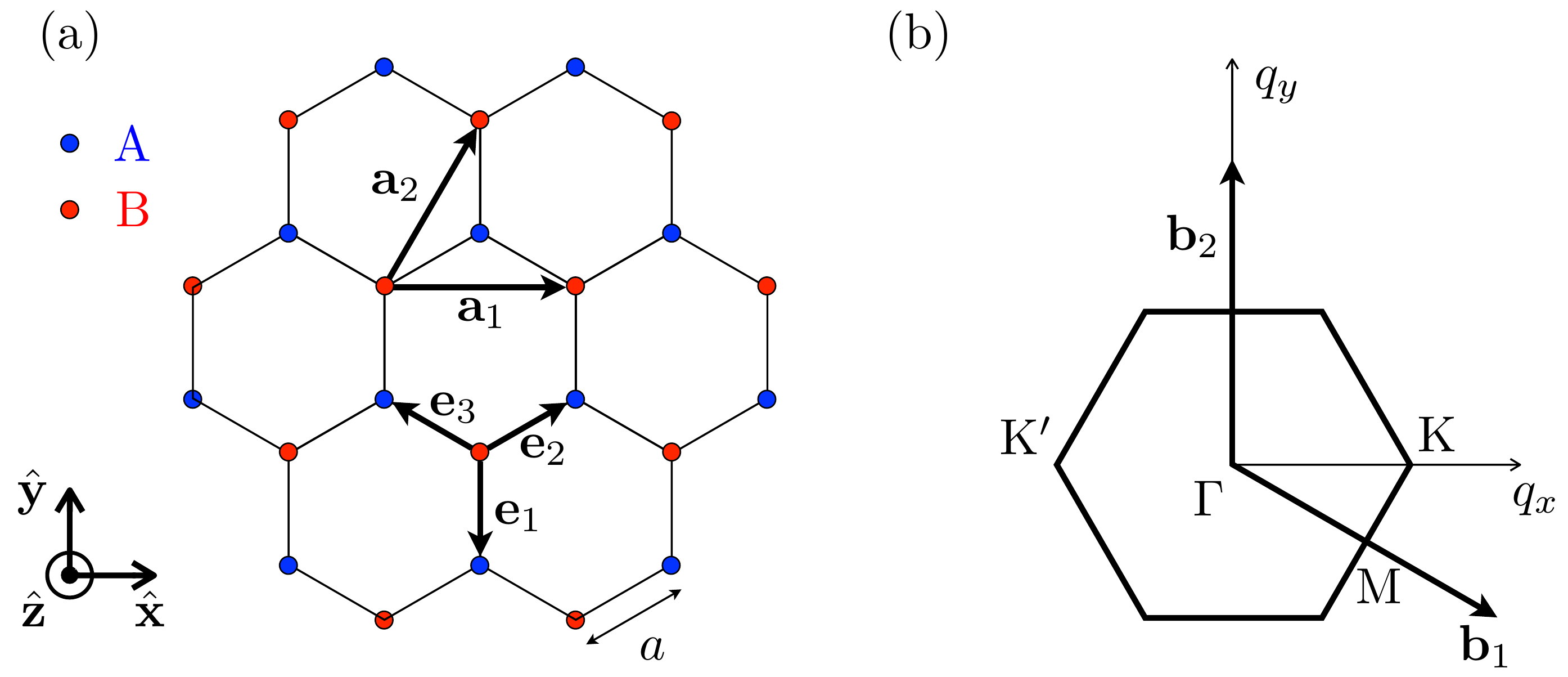}
\caption{\label{fig:lattice}%
(color online). 
(a) Honeycomb lattice with lattice constant $a$ and lattice vectors
$\mathbf{a}_1=a(\sqrt{3},0)$ and 
$\mathbf{a}_2=a(\frac{\sqrt{3}}{2}, \frac 32)$. 
The three vectors 
$\mathbf{e}_1=a(0, -1)$,
$\mathbf{e}_2=a(\frac{\sqrt{3}}{2}, \frac 12)$
and $\mathbf{e}_3=a(-\frac{\sqrt{3}}{2}, \frac 12)$
connect the $\mathrm{A}$ and $\mathrm{B}$ inequivalent lattice sites 
(blue/dark gray and red/light gray dots in the figure).
(b) First Brillouin zone in reciprocal space with primitive vectors 
$\mathbf{b}_1=\frac{2\pi}{3a}(\sqrt{3}, -1)$
and $\mathbf{b}_2=\frac{4\pi}{3a}(0, 1)$.}
\end{figure}

Specifically, we consider an ensemble of identical spherical metallic
nanoparticles of radius $r$ forming a
2D honeycomb lattice with lattice constant $a$ embedded in a dielectric medium with
dielectric constant $\epsilon_\mathrm{m}$ (see Fig.\ \ref{fig:lattice}).
The nanoparticles are located at positions $\mathbf{R}_s$, with $s=\mathrm{A, B}$ a
sublattice index which distinguishes the inequivalent lattice sites.
Each individual
nanoparticle supports a LSP resonance which can be triggered by
an oscillating external electric field with wavelength $\lambda$ much larger than $r$.
Under such a condition, the LSP is a dipolar
collective electronic excitation at
the Mie frequency 
$\omega_0=\omega_\mathrm{p}/\sqrt{1+2\epsilon_\mathrm{m}}$, 
which typically lies in the visible or near-infrared part of the
spectrum 
\cite{mie08_AP}.
Here, $\omega_\mathrm{p}=\sqrt{4\pi n_\mathrm{e}e^2/m_\mathrm{e}}$ is the plasma
frequency, with $n_\mathrm{e}$, $-e$ and $m_\mathrm{e}$ the electron
density, charge, and mass, respectively. 
The LSP corresponding to the electronic center-of-mass excitation can be generally
considered as a quantum bosonic
mode, particularly when the size of the nanoparticle is such that quantization effects are
important 
\cite{kawab66_JPSJ, gerch02_PRA, weick05_PRB,
seoan07_EPJD}. 
The noninteracting part of the
Hamiltonian describing the independent LSPs on the honeycomb lattice sites reads
$H_0=\sum_{s=\mathrm{A, B}}\sum_{\mathbf{R}_s}
[\Pi_s^2(\mathbf{R}_s)/2M+M\omega_0^2h_s^2(\mathbf{R}_s)/2]$,
where $h_s(\mathbf{R})$ is the displacement field associated 
with the electronic center of mass at position $\mathbf{R}$,
$\Pi_s(\mathbf{R})$ its conjugated momentum and
$M=N_\mathrm{e}m_\mathrm{e}$ its mass, with $N_\mathrm{e}$ the number of valence
electrons in each nanoparticle \cite{gerch02_PRA, weick05_PRB}.

The nature of the coupling between LSPs in different nanoparticles depends on
their size and distance. 
Provided that the wavelength associated with each LSP is much larger
than the interparticle distance $a$ 
and that 
$r\lesssim a/3$ \cite{brong00_PRB}, each LSP can be
considered as a point
dipole with dipole moment
$\mathbf{p}=-eN_\mathrm{e}h_s(\mathbf{R})\hat{\mathbf{p}}$
which interacts with the neighboring ones through dipole-dipole
interaction.
Moreover, it has been numerically shown \cite{han09_PRL}
that a quasistatic approximation which only takes into account the
near field
generated by each dipole 
qualitatively reproduces the results of more sophisticated simulations in which
retardation effects are included. Within such a quasistatic approximation,
the interaction between two dipoles $\mathbf{p}$ and $\mathbf{p}'$ located at
$\mathbf{R}$ and $\mathbf{R}'$, respectively, reads 
$\mathcal{V}=
[\mathbf{p}\cdot\mathbf{p}'
-3(\mathbf{p}\cdot\mathbf{n})
(\mathbf{p}'\cdot\mathbf{n})]/
\epsilon_\mathrm{m}|\mathbf{R}-\mathbf{R}'|^3$
with $\mathbf{n}=(\mathbf{R}-\mathbf{R}')/{|\mathbf{R}-\mathbf{R}'|}$
\cite{footnote:dipole}.
In what follows, we assume that in a CP eigenmode all nanoparticles
are polarized in the same direction
$\hat{\mathbf{p}}=\sin{\theta}(\sin{\varphi}\hat{\mathbf{x}}-\cos{\varphi}\hat{\mathbf{y}})+\cos{\theta}\hat{\mathbf{z}}$, 
where $\theta$ is the angle between $\hat{\mathbf{p}}$ and
$\hat{\mathbf{z}}$, and $\varphi$ the angle between the projection of
$\hat{\mathbf{p}}$ in the $xy$ plane and ${\mathbf{e}}_1$ [see Fig.\
\ref{fig:lattice}(a)].
This can be achieved by an external electric field associated with
light of suitable polarization.
We thus write the total Hamiltonian of our system of coupled
LSPs as $H=H_0+H_\mathrm{int}$, 
where the dipole-dipole interaction term reads
\begin{equation}
\label{eq:H_int}
H_\mathrm{int}=\frac{(eN_\mathrm{e})^2}{\epsilon_\mathrm{m}a^3}
\sum_{\mathbf{R}_\mathrm{B}}\sum_{j=1}^3
\mathcal{C}_jh_\mathrm{B}(\mathbf{R}_\mathrm{B})h_\mathrm{A}(\mathbf{R}_\mathrm{B}+\mathbf{e}_j).
\end{equation}
Here,
$\mathcal{C}_j=1-3\sin^2{\theta}\cos^2{(\varphi-2\pi[j-1]/3)}$,
and the vectors $\mathbf{e}_j$ connect the $\mathrm{A}$ and $\mathrm{B}$ sublattices [see Fig.\
\ref{fig:lattice}(a)].
In Eq.\ \eqref{eq:H_int}, we only consider the
dipole-dipole interaction between nearest neighbors, as the effect of
interactions beyond nearest neighbors does not
qualitatively change the plasmonic spectrum \cite{suppl}.

The analogy between
the plasmonic structure of Fig.\ \ref{fig:lattice} and the electronic properties
of graphene becomes transparent by introducing the bosonic ladder operators
$a_\mathbf{R}|b_{\mathbf{R}}=(M\omega_0/2\hbar)^{1/2}h_\mathrm{A|B}(\mathbf{R})
+\mathrm{i}\Pi_\mathrm{A|B}(\mathbf{R})/(2\hbar M\omega_0)^{1/2}$
which satisfy the commutation relations 
$[a^{\phantom\dagger}_\mathbf{R}, a^\dagger_\mathbf{R'}]=
[b^{\phantom\dagger}_\mathbf{R}, b^\dagger_\mathbf{R'}]=\delta_{\mathbf{R},
\mathbf{R'}}$ and $[a^{\phantom\dagger}_\mathbf{R}, b^\dagger_\mathbf{R'}]=0$.
As we will show in the following, the introduction of such operators not only gives access to the plasmon
dispersion (which can be calculated classically as well \cite{suppl}), but also unveils the
Dirac nature of the CP quantum states. 
The harmonic Hamiltonian $H_0$
can be written in terms of the above-mentioned bosonic operators as 
$H_0=\hbar\omega_0
\sum_{\mathbf{R}_\mathrm{A}}a^\dagger_{\mathbf{R}_\mathrm{A}}a^{\phantom\dagger}_{\mathbf{R}_\mathrm{A}}
+\hbar\omega_0
\sum_{\mathbf{R}_\mathrm{B}}b^\dagger_{\mathbf{R}_\mathrm{B}}b^{\phantom\dagger}_{\mathbf{R}_\mathrm{B}}$, 
while Eq.\ \eqref{eq:H_int} transforms into 
\begin{equation}
\label{eq:H_int_bR}
H_\mathrm{int}=
\hbar\Omega\sum_{\mathbf{R}_\mathrm{B}}\sum_{j=1}^3\mathcal{C}_j
b^\dagger_{\mathbf{R}_\mathrm{B}}
\left(a^{\phantom\dagger}_{\mathbf{R}_\mathrm{B}+\mathbf{e}_j}+a^\dagger_{\mathbf{R}_\mathrm{B}+\mathbf{e}_j}\right)
+\mathrm{H.c.}
\end{equation}
In Eq.\ \eqref{eq:H_int_bR}, 
$\Omega=\omega_0(r/a)^3(1+2\epsilon_\mathrm{m})/6\epsilon_\mathrm{m}$,
such that $\Omega\ll\omega_0$.
The first term on the right-hand side of Eq.\ \eqref{eq:H_int_bR} 
resembles the electronic tight-binding Hamiltonian of
graphene \cite{walla47_PR}, except for three major differences: (i) The
Hamiltonian of graphene describes fermionic particles (electrons), while we deal
here with \textit{bosonic} excitations (LSPs). (ii) In graphene, an electron
``hops" from one
lattice site to a neighboring one, i.e., the underlying mechanism linking the
two inequivalent sublattices is purely kinetic. 
In the present case, 
the mechanism coupling the two sublattices is purely
induced by near-field (dipolar) \textit{interactions}, leading to the 
creation of an LSP excitation at lattice site
$\mathbf{R}_\mathrm{B}$ and the annihilation of another LSP at a nearest neighbor
located at $\mathbf{R}_\mathrm{B}+\mathbf{e}_j$.
(iii) In (unstrained) graphene, the hopping matrix element between two neighboring atoms is
the same for all three bonds. In contrast, in our case the three energy scales
$\hbar\Omega\mathcal{C}_{j}$ in Eq.\
\eqref{eq:H_int} are in general different 
and can be tuned by the direction of the 
polarization $\hat{\mathbf{p}}$ of the CP eigenmode, which can be
controlled by means of an external light field.
For $0<\theta\leqslant\theta_0$ and
$\pi-\theta_0\leqslant\theta<\pi$ with $\theta_0=\arcsin{\sqrt{1/3}}$,
the coefficients $\mathcal{C}_j$ are all positive for any $\varphi$ and have
different values, resulting in different couplings between the bonds, thus mimicking
the effect of strain in the lattice \cite{suzuu02_PRB}. 
For $\theta_0<\theta<\pi-\theta_0$, the signs of the coefficients
$\mathcal{C}_j$ depend on $\varphi$, and the analogy with strained graphene is
no longer valid. In the special case where
$\mathcal{C}_1=\mathcal{C}_2=\mathcal{C}_3$ (for $\theta=0$ or $\pi$), we expect
the CP spectrum to resemble that of the electronic band
structure in graphene, since the Bloch
theorem does not depend on the quantum statistics of the particles one
considers, but only on the structure of the periodic lattice.  
This fact is also responsible for the conical dispersion presented by other
systems with honeycomb symmetry \cite{halda08_PRL, torre12_PRL, tarru12_Nature}.
As we will
now show, two slight differences appear in the CP dispersion as
compared to the graphene band structure, 
i.e., the effect of the Hamiltonian $H_0$ 
is to produce a global energy shift (by an amount $\hbar\omega_0$), while the ``anomalous" term
$\propto b^\dagger_{\mathbf{R}_\mathrm{B}}a^\dagger_{\mathbf{R}_\mathrm{B}+\mathbf{e}_j}$ in 
Eq.\ \eqref{eq:H_int_bR} introduces corrections of
order $(\Omega/\omega_0)^2$ to the spectrum.

Introducing the bosonic operators in
momentum space $a_\mathbf{q}$ and $b_\mathbf{q}$ through 
$a_{\mathbf{R}}|b_{\mathbf{R}}=
\mathcal{N}^{-1/2}\sum_{\mathbf{q}}\exp{(\mathrm{i}\mathbf{q}\cdot\mathbf{R})}a_\mathbf{q}|b_\mathbf{q}$,
with $\mathcal{N}$ the number of unit cells of the honeycomb lattice, 
the Hamiltonian $H=H_0+H_\mathrm{int}$ transforms into
$H=\hbar\omega_0\sum_\mathbf{q}
(a^\dagger_\mathbf{q}a^{\phantom\dagger}_\mathbf{q}
+b^\dagger_\mathbf{q}b^{\phantom\dagger}_\mathbf{q})
+\hbar\Omega\sum_\mathbf{q}
[f^{\phantom *}_\mathbf{q} b^\dagger_\mathbf{q}(a^{\phantom\dagger}_\mathbf{q}+a^\dagger_{-\mathbf{q}})
+\mathrm{H.c.}]$
with 
$f_\mathbf{q}=\sum_{j=1}^3\mathcal{C}_j\exp{(\mathrm{i}\mathbf{q}\cdot\mathbf{e}_j)}$.
The latter Hamiltonian 
is diagonalized by two successive Bogoliubov
transformations. 
First, we introduce the two bosonic operators 
$\alpha_\mathbf{q}^\pm=[(f_\mathbf{q}/|f_\mathbf{q}|)a_\mathbf{q}\pm
b_\mathbf{q}]/\sqrt{2}$
in terms of which we obtain
$H=
\sum_{\tau=\pm}\sum_{\mathbf{q}}
[
(\hbar\omega_0+\tau\hbar\Omega|f_\mathbf{q}|){\alpha_\mathbf{q}^\tau}^\dagger{\alpha_\mathbf{q}^\tau}
+\tau\frac{\hbar\Omega|f_\mathbf{q}|}{2}
({\alpha_\mathbf{q}^\tau}^\dagger{\alpha_\mathbf{-q}^{\tau}}^\dagger+\mathrm{H.c.})
  ]$.
Second, we define two new bosonic modes 
$\beta_\mathbf{q}^\pm=\cosh{\vartheta_\mathbf{q}^\pm}\alpha_\mathbf{q}^\pm
-\sinh{\vartheta_\mathbf{q}^\pm}{\alpha_\mathbf{-q}^\pm}^\dagger$, 
with 
$\cosh{\vartheta_\mathbf{q}^\pm}=
2^{-1/2}[
{(1\pm\Omega|f_\mathbf{q}|/\omega_0)}/{{(1\pm2\Omega|f_\mathbf{q}|/\omega_0)^{1/2}}}
+1
]^{1/2}$
and
$\sinh{\vartheta_\mathbf{q}^\pm}=\mp
2^{-1/2}[
{(1\pm\Omega|f_\mathbf{q}|/\omega_0)}/{{(1\pm2\Omega|f_\mathbf{q}|/\omega_0)^{1/2}}}
-1
]^{1/2}$, 
which
diagonalize the Hamiltonian $H$ 
as
\begin{equation}
\label{eq:dispersion}
H=
\sum_{\tau=\pm}\sum_{\mathbf{q}}
\hbar\omega_\mathbf{q}^\tau{\beta_\mathbf{q}^\tau}^\dagger\beta_\mathbf{q}^\tau,
\quad
\omega_\mathbf{q}^\pm=\omega_0\sqrt{1\pm2\frac{\Omega}{\omega_0}|f_\mathbf{q}|}.
\end{equation}
The two CP branches 
reduce to 
$\omega_\mathbf{q}^\pm\simeq\omega_0\pm\Omega|f_\mathbf{q}|$ to first order in 
$\Omega/\omega_0\ll1$, for which we have
$\beta_\mathbf{q}^\pm\simeq\alpha_\mathbf{q}^\pm$ \cite{suppl}.

\begin{figure*}[tb]
\includegraphics[width=\linewidth]{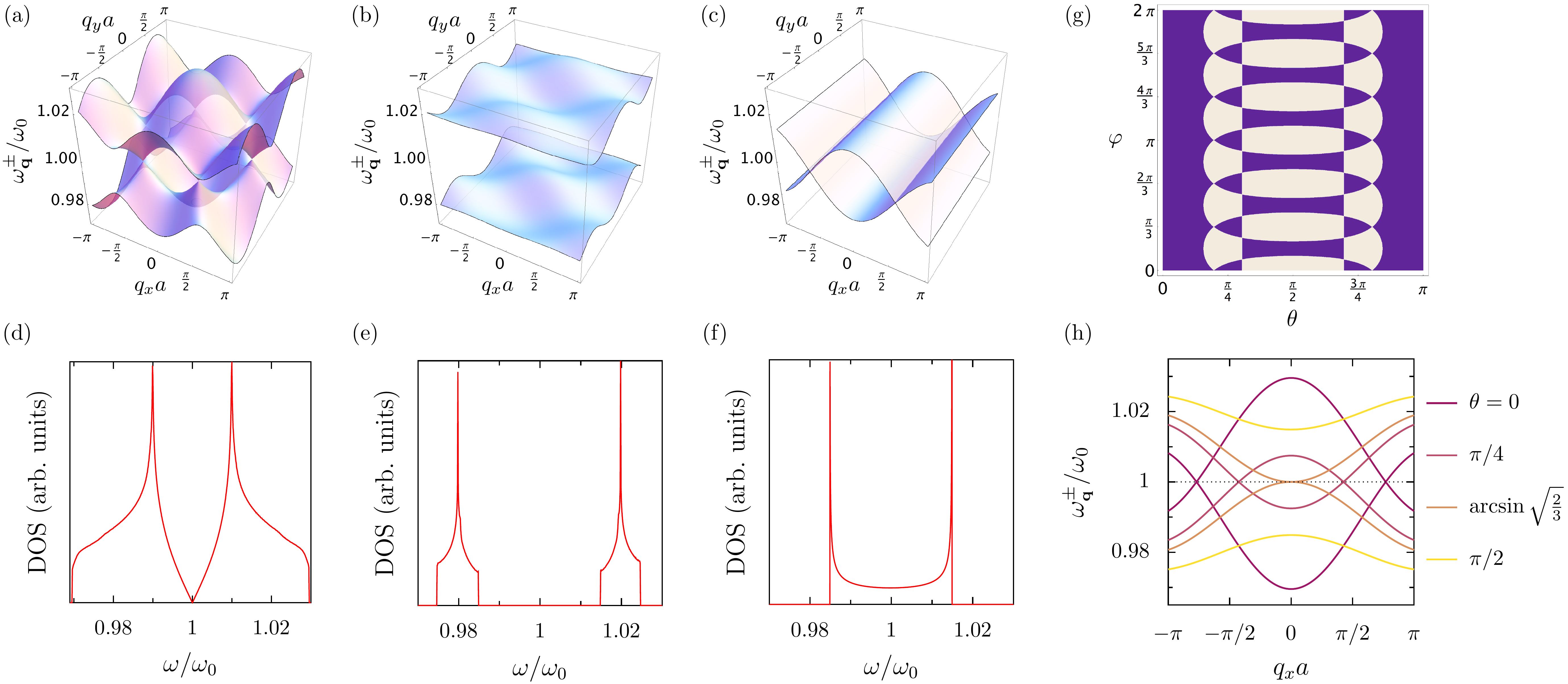}
\caption{\label{fig:dispersion}%
(color online). 
(a)--(c) Collective plasmon dispersion relation from Eq.\ \eqref{eq:dispersion}
and (d)--(f) corresponding density of states for (a),(d) the out-of-plane 
mode ($\theta=0$), (b),(e) one in-plane mode
($\theta=\pi/2$), and (c),(f) $\theta=\arcsin{\sqrt{1/3}}$. 
(g) Polarization angles $(\theta, \varphi)$ for which the collective plasmon dispersion is gapless (dark blue
regions) and gapped (white regions).
(h) Collective plasmon dispersion along the
$\mathrm{K}'\Gamma\mathrm{K}$ direction ($q_y=0$) for different orientations
$\theta$ of the dipoles.
In the figure, $\varphi=0$ and $\Omega/\omega_0=0.01$.}
\end{figure*}

The dispersion in Eq.\ \eqref{eq:dispersion} is shown in Fig.\ \ref{fig:dispersion}
in the case of a polarization $\hat{\mathbf{p}}$ perpendicular to the plane of
the honeycomb lattice [$\theta=0$, Fig.\ \ref{fig:dispersion}(a)], in the case of
an in-plane polarization [$\theta=\pi/2$, $\varphi=0$, Fig.\
\ref{fig:dispersion}(b)], and for the special case
$\theta=\arcsin{\sqrt{1/3}}$, $\varphi=0$ [Fig.\ \ref{fig:dispersion}(c)].
In the first case [Fig.\ \ref{fig:dispersion}(a)], we have gapless modes with 
two inequivalent Dirac cones centered at the $\mathrm{K}$ and $\mathrm{K}'$ points 
located at $\pm\mathbf{K}=\frac{4\pi}{3\sqrt{3}a}(\pm1, 0)$ in the first Brillouin zone
[cf.\  Fig.\ \ref{fig:lattice}(b)], while in the second case, the modes are gapped [Fig.\ \ref{fig:dispersion}(b)].
The dispersion shown in Fig.\ \ref{fig:dispersion}(c) corresponds to a
polarization for which $\mathcal{C}_1=0$ in Eq.\ \eqref{eq:H_int}, i.e., the
bonds linked by $\mathbf{e}_1$ [cf.\ Fig.\ \ref{fig:lattice}(a)] are ineffective
and the system is effectively translationally invariant along one direction. Hence, the CP dispersion in Fig.\
\ref{fig:dispersion}(c) does not
depend on $q_y$ and presents Dirac ``lines".

The analogy between the dispersion shown in Fig.\ \ref{fig:dispersion}(a) and
the electronic band structure of graphene \cite{walla47_PR} is striking.
Close to the two inequivalent Dirac points $\mathrm{K}$ and $\mathrm{K}'$ 
[see Fig.\ \ref{fig:lattice}(b)], 
the function $f_\mathbf{q}$ expands as 
$f_\mathbf{q}\simeq-\frac{3a}{2}(\pm k_x+\mathrm{i}k_y)$ with
$\mathbf{q}=\pm\mathbf{K}+\mathbf{k}$ ($|\mathbf{k}|\ll|\mathbf{K}|$), such that 
the dispersion in Eq.\ \eqref{eq:dispersion} is linear and forms a Dirac cone,  
$\omega_\mathbf{k}^\pm\simeq\omega_0\pm v|\mathbf{k}|$, with
group velocity $v=3\Omega a/2$. This feature is consistent with numerical
analysis \cite{han09_PRL}.
Moreover, by expanding Eq.\ \eqref{eq:dispersion} in the vicinity of the Dirac points, we
can identify the Hamiltonian 
$H^\mathrm{eff}=\sum_{\mathbf{k}}\hat{\Psi}^\dagger_\mathbf{k}\mathcal{H}_\mathbf{k}^\mathrm{eff}\hat{\Psi}^{\phantom\dagger}_\mathbf{k}$
that effectively describes the CPs.
Here 
$\hat{\Psi}_\mathbf{k}=(a_{\mathbf{k},\mathrm{K}}, b_{\mathbf{k},\mathrm{K}},
b_{\mathbf{k},\mathrm{K'}}, a_{\mathbf{k},\mathrm{K'}})$ is a spinor operator, 
where $\mathrm{K}$ and $\mathrm{K}'$ denote the valley indices associated with 
the inequivalent Dirac points, and the $4\times4$ Hamiltonian reads
\begin{equation}
\label{eq:H_eff}
\mathcal{H}_\mathbf{k}^\mathrm{eff}=\hbar\omega_0\mathbb{1}-\hbar v\tau_z\otimes\boldsymbol{\sigma}\cdot\mathbf{k}.
\end{equation}
In this notation, $\mathbb{1}$ corresponds to the identity matrix,
$\tau_z$ to the Pauli matrix acting on the valley space ($\mathrm{K}$/$\mathrm{K'}$),
while $\boldsymbol{\sigma}=(\sigma_x, \sigma_y)$ is the vector of Pauli matrices acting on 
the sublattice space ($\mathrm{A}$/$\mathrm{B}$).
Up to a global energy shift of $\hbar\omega_0$, Eq.\ \eqref{eq:H_eff}
corresponds to a \textit{massless Dirac Hamiltonian} that is fulfilled by CPs,
in complete analogy with electrons in graphene \cite{walla47_PR}.  
The CP eigenstates of Eq.\ \eqref{eq:H_eff},
$\psi_{\mathbf{k},\mathrm{K}}^\pm=\frac{1}{\sqrt{2}}(1, \mp \mathrm{e}^{\mathrm{i}\xi_\mathbf{k}}, 0, 0)$ 
and 
$\psi_{\mathbf{k},\mathrm{K'}}^\pm=\frac{1}{\sqrt{2}}(0, 0, 1, \pm
\mathrm{e}^{\mathrm{i}\xi_\mathbf{k}})$ with $\xi_\mathbf{k}=\arctan{(k_y/k_x)}$,
are characterized by chirality 
$\boldsymbol{\sigma}\cdot\hat{\mathbf{k}}=\pm\mathbb{1}$. As
a consequence, CPs will show similar effects to electrons in graphene like
a Berry phase of $\pi$ \cite{novos05_Nature} and the absence of backscattering off smooth inhomogeneities
\cite{cheia06_PRB}.
This could have crucial implications for the efficient plasmonic propagation in
array-based metamaterials.

In Fig.\ \ref{fig:dispersion}, the panels (d)--(f) show the DOS corresponding to
the spectrum illustrated in the panels (a)--(c). It is interesting to notice the
tunability of the DOS with the direction of the polarization, as well as the
emergence of van Hove singularities.
The latter are associated with Lifshitz
transitions \cite{lifsh60_JETP} in the topology of equipotential lines that percolate at specific
energies. The tunability of van Hove singularities in the spectrum could be of
crucial importance to increase the coupling of light of different wavelengths
with extended CP modes.

For an arbitrary polarization of the LSPs, we can determine if the CP
dispersion is gapless by imposing $|f_\mathbf{q}|=0$ in Eq.\
\eqref{eq:dispersion}, which leads to the
condition 
$0\leqslant[(\mathcal{C}_2+\mathcal{C}_3)^2-\mathcal{C}_1^2]/4\mathcal{C}_2\mathcal{C}_3\leqslant1$
for having gapless plasmonic modes \cite{suppl}. 
In Fig.\ \ref{fig:dispersion}(g), we show in dark blue the 
regions of stability of a massless Dirac spectrum in the 
$(\theta, \varphi)$ parameter space
for which one has gapless plasmon modes, an example of
which is shown in Fig.\ \ref{fig:dispersion}(a). In Fig.\ \ref{fig:dispersion}(g), the
white regions correspond to polarizations for which the CP dispersion is
gapped [as an example, see Fig.\ \ref{fig:dispersion}(b)].
Thus, changing the
polarization allows one to \textit{qualitatively} change the CP 
spectrum. 
This is further illustrated in Fig.\ \ref{fig:dispersion}(h) where we show
the CP dispersion along the
$\mathrm{K}'\Gamma\mathrm{K}$ direction [see Fig.\ \ref{fig:lattice}(b)] for
different angles $\theta$ of the polarization (in the figure, $\varphi=0$).
As one can see from Fig.\ \ref{fig:dispersion}(h), the two inequivalent Dirac points
located at $\mathrm{K}$ and $\mathrm{K}'$ for $\theta=0$ drift as one increases
$\theta$ and they merge at $\mathbf{q}=0$ for $\theta=\arcsin{\sqrt{2/3}}$, forming 
parabolic bands, to finally open a gap for $\theta>\arcsin{\sqrt{2/3}}$
(exemplified by $\theta=\pi/2$ in the figure).

A limitation on the experimental observability of the CP dispersion is 
plasmonic damping, which tends to blur the resonance frequencies. 
In order to estimate the feasibility of such experiments, we compare
the bandwidth of the CP dispersion to the losses in individual nanoparticles. 
In the latter, two main sources of
dissipation arise: (i) radiation damping with decay rate
$\gamma_\mathrm{rad}=2r^3\omega_0^4/3c^3$ \cite{crowe68_PR} ($c$ is
the speed of light) which dominates for larger nanoparticle sizes, and 
(ii) Landau damping with decay rate $\gamma_\mathrm{L}=3v_\mathrm{F}g/4r$
\cite{kawab66_JPSJ, weick05_PRB, mie08_AP} 
($v_\mathrm{F}$ is the bulk Fermi
velocity and $g$ a constant of the order of one)
which dominates for smaller sizes.
Hence, there exists an optimal size
$r_\mathrm{opt}=(3v_\mathrm{F}gc^3/8)^{1/4}/\omega_0$ for which the total damping
$\gamma_\mathrm{tot}=\gamma_\mathrm{rad}+\gamma_\mathrm{L}$ is minimal. For Ag
nanoparticles, we find $r_\mathrm{opt}=\unit[8]{nm}$ for which
$\gamma_\mathrm{tot}=\unit[0.1]{eV}/\hbar$.
With an interparticle distance $a=3r_\mathrm{opt}$ which maximizes the dipolar coupling
between nanoparticles \cite{brong00_PRB}, we find that the bandwidth 
is of the order of
$\Delta\omega=\omega_0^+-\omega_0^-=\omega_0/9=\unit[0.6]{eV}/\hbar$ at the
center of the Brillouin zone 
(for $\epsilon_\mathrm{m}=1$, and for the out-of-plane polarization). Thus $\Delta\omega$ is sufficiently large when compared to
$\gamma_\mathrm{tot}$ that the plasmon excitation is well defined and hence clearly
measurable. 
Moreover, the appropriate use of active (gain-enhanced) media \cite{hess12_NatureMat}
might increase the observability of the CP dispersion.

A last comment is in order about the excitation of CPs by external photons, whose in-plane
momentum must match the plasmonic one. In fact, the vicinity of the Dirac points typically
lies outside the light cone.
In order to overcome this momentum mismatch and observe the Dirac plasmons, one might
add an extra periodic modulation of the lattice
to allow grating coupling between the incident
light and the desired collective modes \cite{maier}.
Another alternative might be to use a non-linear technique to overcome the momentum
mismatch \cite{renge09_PRL}.

In conclusion, we demonstrated the strong analogies between
the physical properties of electrons in graphene and those of collective plasmon modes in a 2D
honeycomb lattice of metallic nanoparticles. 
Whereas the electronic states of
graphene can be described by massless Dirac fermions, the
CP eigenstates correspond to massless Dirac-like bosonic excitations. The
spectrum of the latter can be fully tuned by the polarization of an external
light field, opening exciting new possibilities for controlling the propagation of
electromagnetic radiation with subwavelength lateral confinement in plasmonic
metamaterials.

We thank C.\ Gorini, R.\ A.\ 
Jalabert, and D.\ Weinmann for stimulating discussions.
O.H.\ would like to acknowledge support by the Leverhulme Trust. 
E.M.\ acknowledges the support of the Royal Society via the research grant
RG110429.


\onecolumngrid
\newpage
\begin{center}
{\large\textbf{Supplemental Material}}
\end{center}

\setcounter{equation}{0}
\setcounter{figure}{0}
\renewcommand{\theequation}{S\arabic{equation}}
\renewcommand{\thefigure}{S\arabic{figure}}
\renewcommand\figurename{Supplementary Figure}

\section{Classical dynamics of the collective plasmon}
In what follows, we derive the collective plasmon dispersion from classical equations
of motion, and show that one recovers the dispersion obtained quantum
mechanically in the main text. Moreover, we show that the phase
relations linking the amplitudes of motion on the $\mathrm{A}$ and $\mathrm{B}$
sublattices are the same as for the eigenspinors in the quantum case.

Considering only the dipole-dipole interaction between nearest neighbors on the
honeycomb lattice, the Hamiltonian of the system reads 
(see the main text for a definition of the different terms)
\begin{equation}
\label{eq:H_SM}
H=
\sum_{s=\mathrm{A, B}}\sum_{\mathbf{R}_s}
\left[\frac{\Pi_s^2(\mathbf{R}_s)}{2M}+\frac{M}{2}\omega_0^2h_s^2(\mathbf{R}_s)\right]
+
\frac{(eN_\mathrm{e})^2}{\epsilon_\mathrm{m}a^3}
\sum_{\mathbf{R}_\mathrm{B}}\sum_{j=1}^3\mathcal{C}_j
h_\mathrm{B}(\mathbf{R}_\mathrm{B})h_\mathrm{A}(\mathbf{R}_\mathrm{B}+\mathbf{e}_j).
\end{equation}
The corresponding equations of motion, 
$\dot h_s(\mathbf{R})=\partial H/\partial \Pi_s(\mathbf{R})$, 
$\dot \Pi_s(\mathbf{R})=-\partial H/\partial h_s(\mathbf{R})$, 
$s=\mathrm{A, B}$, reduce to 
\begin{subequations}
\label{eq:system_nn}
\begin{align}
\ddot h_\mathrm{A}(\mathbf{R})+\omega_0^2h_\mathrm{A}(\mathbf{R})&=-2\omega_0\Omega
\sum_{j=1}^3\mathcal{C}_j h_\mathrm{B}(\mathbf{R}-\mathbf{e}_j),\\
\ddot h_\mathrm{B}(\mathbf{R})+\omega_0^2h_\mathrm{B}(\mathbf{R})&=-2\omega_0\Omega
\sum_{j=1}^3\mathcal{C}_j h_\mathrm{A}(\mathbf{R}+\mathbf{e}_j),
\end{align}
\end{subequations}
with
$\Omega=\omega_0(r/a)^3(1+2\epsilon_\mathrm{m})/6\epsilon_\mathrm{m}\ll\omega_0$.
Introducing the Fourier decomposition 
\begin{equation}
\label{eq:Fourier}
h_s(\mathbf{R})=\frac{1}{\sqrt{\mathcal{N}}}\sum_{\mathbf{q}}
\mathrm{e}^{\mathrm{i}(\mathbf{q}\cdot\mathbf{R}-\omega t)}\tilde h_s(\mathbf{q}),
\quad
s=\mathrm{A, B}, 
\end{equation}
where $\mathcal{N}$ is the number of unit cells, 
the system of equations \eqref{eq:system_nn} transforms into
\begin{equation}
\label{eq:matrix}
\begin{pmatrix}
\omega_0^2-\omega^2 & 2\omega_0\Omega f_{\mathbf{q}}^*
\vspace{.2truecm}\\
2\omega_0\Omega f_{\mathbf{q}} & \omega_0^2-\omega^2
\end{pmatrix}
\begin{pmatrix}
\tilde h_\mathrm{A}(\mathbf{q})
\vspace{.2truecm}\\
\tilde h_\mathrm{B}(\mathbf{q})
\end{pmatrix}
=0, 
\end{equation}
where
$f_\mathbf{q}=\sum_{j=1}^3\mathcal{C}_j\exp{(\mathrm{i}\mathbf{q}\cdot\mathbf{e}_j)}$.
The above system of equations only has nontrivial solutions when
$\omega=\omega_\mathbf{q}^\pm$, where 
\begin{subequations}
\label{eq:dispersion_SM}
\begin{align}
\omega_\mathbf{q}^\pm&=\omega_0\sqrt{1\pm2\frac{\Omega}{\omega_0}|f_\mathbf{q}|}
\\
&\simeq\omega_0\pm\Omega|f_\mathbf{q}|.
\end{align}
\end{subequations}
With Eq.\ \eqref{eq:dispersion_SM}, we thus recover the collective plasmon dispersion derived
quantum mechanically in the main text.

As mentioned in the main text, the dispersion \eqref{eq:dispersion_SM}
resembles the one of electrons in graphene \cite{walla47_PR_SM, castr09_RMP_SM} 
when the polarization of the dipoles points perpendicular
to the honeycomb lattice. For wave vectors $\mathbf{k}$ in the vicinity of the two Dirac
points $\mathrm{K}$ and $\mathrm{K'}$, the dispersion \eqref{eq:dispersion_SM} forms a cone,
$\omega^\pm_\mathbf{k}=\omega_0\pm v|\mathbf{k}|$. 
In that case, one obtains from Eq.\ \eqref{eq:matrix} with
$\omega=\omega_\mathbf{q}^\pm$ that the amplitudes $\tilde
h_{\mathrm{A},\mathrm{K}}^\pm(\mathbf{k})$ and
$\tilde h_{\mathrm{B},\mathrm{K}}^\pm(\mathbf{k})$ corresponding to the $+$ and $-$
branches close to the $\mathrm{K}$ point are linked by 
$\tilde h_{\mathrm{B},\mathrm{K}}^\pm(\mathbf{k})/\tilde
h_{\mathrm{A},\mathrm{K}}^\pm(\mathbf{k})=\mp\mathrm{e}^{\mathrm{i}\xi_\mathbf{k}}$, 
while close to the $\mathrm{K}'$ point, one has 
$\tilde h_{\mathrm{A},\mathrm{K'}}^\pm(\mathbf{k})/\tilde
h_{\mathrm{B},\mathrm{K'}}^\pm(\mathbf{k})=\pm\mathrm{e}^{\mathrm{i}\xi_\mathbf{k}}$, 
with $\xi_\mathbf{k}=\arctan{(k_y/k_x)}$. Interestingly, these phase relations linking
the $\mathrm{A}$ and $\mathrm{B}$ sublattices are the
same as for the eigenspinors 
$\psi^\pm_{\mathbf{k},\mathrm{K}}=\frac{1}{\sqrt{2}}(1, \mp\mathrm{e}^{\mathrm{i}\xi_\mathbf{k}}, 0, 0)$ and
$\psi^\pm_{\mathbf{k},\mathrm{K'}}=\frac{1}{\sqrt{2}}(0, 0, 1, \pm\mathrm{e}^{\mathrm{i}\xi_\mathbf{k}})$
in the quantum case.

\section{Plasmon dispersion with dipole-dipole interaction beyond nearest neighbors}
In the main text, we only consider the interaction between nearest neighbors on
the honeycomb lattice. However, as the dipole-dipole interaction decays as one
over the cube of the interparticle distance, we must check the robustness of our
results against the effect of interactions beyond nearest neighbors. In the
following, we show that the plasmon dispersion \eqref{eq:dispersion_SM} is only slightly modified by
interactions beyond nearest neighbors, and that the interaction between the
nearest neighbors alone captures the relevant physics of the problem.

\subsection{Next nearest neighbors}
When considering the dipole-dipole interaction between next nearest neighbors 
in addition to that of nearest neighbors, the Hamiltonian \eqref{eq:H_SM} has to be
supplemented with the interaction term 
\begin{equation}
\label{eq:H_int_2}
H_\mathrm{int}^{(2)}=
\frac{(eN_\mathrm{e})^2}{3\sqrt{3}\epsilon_\mathrm{m}a^3}
\sum_{s=\mathrm{A, B}}\sum_{\mathbf{R}_s}\sum_{j=1}^3\mathcal{S}_j
h_s(\mathbf{R}_s)h_s(\mathbf{R}_s+\mathbf{e}_j^{(2)}), 
\end{equation}
where the vectors $\mathbf{e}_1^{(2)}=\mathbf{a}_1$, 
$\mathbf{e}_2^{(2)}=\mathbf{a}_2-\mathbf{a}_1$ and 
$\mathbf{e}_3^{(2)}=-\mathbf{a}_2$ connect the next nearest 
neighbors on the honeycomb lattice, with $\mathbf{a}_1$ and $\mathbf{a}_2$ the
lattice vectors (see Fig.\ 1 in the main text). 
In Eq.\ \eqref{eq:H_int_2}, 
$\mathcal{S}_j=1-3\sin^2{\theta}\sin^2{(\varphi-2\pi[j-1]/3)}$.
With Eqs.\ \eqref{eq:H_SM} and \eqref{eq:H_int_2}, the equations of
motion read
\begin{subequations}
\begin{align}
\ddot h_\mathrm{A}(\mathbf{R})+\omega_0^2h_\mathrm{A}(\mathbf{R})&=-2\omega_0\Omega
\sum_{j=1}^3\mathcal{C}_j h_\mathrm{B}(\mathbf{R}-\mathbf{e}_j)
-\frac{2\omega_0\Omega}{3\sqrt{3}}
\sum_{j=1}^3\mathcal{S}_j[h_\mathrm{A}(\mathbf{R}+\mathbf{e}_j^{(2)})
+h_\mathrm{A}(\mathbf{R}-\mathbf{e}_j^{(2)})]
,\\
\ddot h_\mathrm{B}(\mathbf{R})+\omega_0^2h_\mathrm{B}(\mathbf{R})&=-2\omega_0\Omega
\sum_{j=1}^3\mathcal{C}_j h_\mathrm{A}(\mathbf{R}+\mathbf{e}_j)
-\frac{2\omega_0\Omega}{3\sqrt{3}}
\sum_{j=1}^3\mathcal{S}_j[h_\mathrm{B}(\mathbf{R}+\mathbf{e}_j^{(2)})
+h_\mathrm{B}(\mathbf{R}-\mathbf{e}_j^{(2)})].
\end{align}
\end{subequations}
Together with the decomposition \eqref{eq:Fourier}, we obtain the linear system 
\begin{equation}
\begin{pmatrix}
\omega_0^2-\omega^2+\dfrac{4\omega_0\Omega}{3\sqrt{3}}\mathrm{Re}f_{\mathbf{q}}^{(2)}
& 2\omega_0\Omega f_{\mathbf{q}}^*
\vspace{.2truecm}\\
2\omega_0\Omega f_{\mathbf{q}} 
& \omega_0^2-\omega^2+\dfrac{4\omega_0\Omega}{3\sqrt{3}}\mathrm{Re}f_{\mathbf{q}}^{(2)}
\end{pmatrix}
\begin{pmatrix}
\tilde h_\mathrm{A}(\mathbf{q})
\vspace{.6truecm}\\
\tilde h_\mathrm{B}(\mathbf{q})
\end{pmatrix}
=0 
\end{equation}
which has nontrivial solutions only if $\omega=\omega_\mathbf{q}^{\pm(2)}$, with 
\begin{subequations}
\label{eq:dispersion_2}
\begin{align}
\omega_\mathbf{q}^{\pm(2)}&=\omega_0\sqrt{1\pm2\frac{\Omega}{\omega_0}|f_\mathbf{q}|
+\frac{4}{3\sqrt{3}}\frac{\Omega}{\omega_0}\mathrm{Re}f_\mathbf{q}^{(2)}}
\\
&\simeq\omega_\mathbf{q}^{\pm}+\frac{2}{3\sqrt{3}}\frac{\Omega}{\omega_0}\mathrm{Re}f_\mathbf{q}^{(2)},
\end{align}
\end{subequations}
where $\omega_\mathbf{q}^{\pm}$ is defined in Eq.\ \eqref{eq:dispersion_SM}.
Here, we defined  
$f_\mathbf{q}^{(2)}=\sum_{j=1}^3\mathcal{S}_j\exp{(\mathrm{i}\mathbf{q}\cdot\mathbf{e}_j^{(2)})}$.

\subsection{Third nearest neighbors}
The Hamiltonian corresponding to the dipole-dipole interaction between third
nearest neighbors reads
\begin{equation}
\label{eq:H_int_3}
H_\mathrm{int}^{(3)}=\frac{(eN_\mathrm{e})^2}{8\epsilon_\mathrm{m}a^3}
\sum_{\mathbf{R}_\mathrm{B}}\sum_{j=1}^3\mathcal{C}_j
h_\mathrm{B}(\mathbf{R}_\mathrm{B})h_\mathrm{A}(\mathbf{R}_\mathrm{B}-2\mathbf{e}_j).
\end{equation}
With Eqs.\ \eqref{eq:H_SM}, \eqref{eq:H_int_2} and \eqref{eq:H_int_3}, the
equations of motion now read
\begin{subequations}
\begin{align}
\ddot h_\mathrm{A}(\mathbf{R})+\omega_0^2h_\mathrm{A}(\mathbf{R})&=-2\omega_0\Omega
\sum_{j=1}^3\mathcal{C}_j \left[h_\mathrm{B}(\mathbf{R}-\mathbf{e}_j)+\frac{1}{8}h_\mathrm{B}(\mathbf{R}+2\mathbf{e}_j)\right]
-\frac{2\omega_0\Omega}{3\sqrt{3}}
\sum_{j=1}^3\mathcal{S}_j[h_\mathrm{A}(\mathbf{R}+\mathbf{e}_j^{(2)})
+h_\mathrm{A}(\mathbf{R}-\mathbf{e}_j^{(2)})]
,\\
\ddot h_\mathrm{B}(\mathbf{R})+\omega_0^2h_\mathrm{B}(\mathbf{R})&=-2\omega_0\Omega
\sum_{j=1}^3\mathcal{C}_j
\left[h_\mathrm{A}(\mathbf{R}+\mathbf{e}_j)+\frac{1}{8}h_\mathrm{A}(\mathbf{R}-2\mathbf{e}_j)\right]
-\frac{2\omega_0\Omega}{3\sqrt{3}}
\sum_{j=1}^3\mathcal{S}_j[h_\mathrm{B}(\mathbf{R}+\mathbf{e}_j^{(2)})
+h_\mathrm{B}(\mathbf{R}-\mathbf{e}_j^{(2)})].
\end{align}
\end{subequations}
With Eq.\ \eqref{eq:Fourier}, such a system reduces to
\begin{equation}
\begin{pmatrix}
\omega_0^2-\omega^2+\dfrac{4\omega_0\Omega}{3\sqrt{3}}\mathrm{Re}f_{\mathbf{q}}^{(2)}
& 2\omega_0\Omega f_{\mathbf{q}}^*+\dfrac{\omega_0\Omega}{4}f_{2\mathbf{q}}
\vspace{.2truecm}\\
2\omega_0\Omega f_{\mathbf{q}}+\dfrac{\omega_0\Omega}{4}f_{2\mathbf{q}}^*
& \omega_0^2-\omega^2+\dfrac{4\omega_0\Omega}{3\sqrt{3}}\mathrm{Re}f_{\mathbf{q}}^{(2)}
\end{pmatrix}
\begin{pmatrix}
\tilde h_\mathrm{A}(\mathbf{q})
\vspace{.6truecm}\\
\tilde h_\mathrm{B}(\mathbf{q})
\end{pmatrix}
=0 
\end{equation}
which only has nontrivial solutions if $\omega=\omega_\mathbf{q}^{\pm(3)}$, 
with 
\begin{subequations}
\label{eq:dispersion_3}
\begin{align}
\omega_\mathbf{q}^{\pm(3)}&=\omega_0\sqrt{1\pm2\frac{\Omega}{\omega_0}
\sqrt{|f_\mathbf{q}|^2+\frac 14 \mathrm{Re}(f_\mathbf{q}f_{2\mathbf{q}})
+\frac{1}{64}|f_{2\mathbf{q}}|^2}
+\frac{4}{3\sqrt{3}}\frac{\Omega}{\omega_0}\mathrm{Re}f_\mathbf{q}^{(2)}}
\\
&\simeq\omega_\mathbf{q}^{\pm(2)}
\pm\Omega\left[
\sqrt{|f_\mathbf{q}|^2+\frac 14 \mathrm{Re}(f_\mathbf{q}f_{2\mathbf{q}})
+\frac{1}{64}|f_{2\mathbf{q}}|^2}-|f_\mathbf{q}|
\right],
\end{align}
\end{subequations}
with $\omega_\mathbf{q}^{\pm(2)}$ defined in Eq.\ \eqref{eq:dispersion_2}.

\begin{figure}[tb]
\includegraphics[width=8truecm]{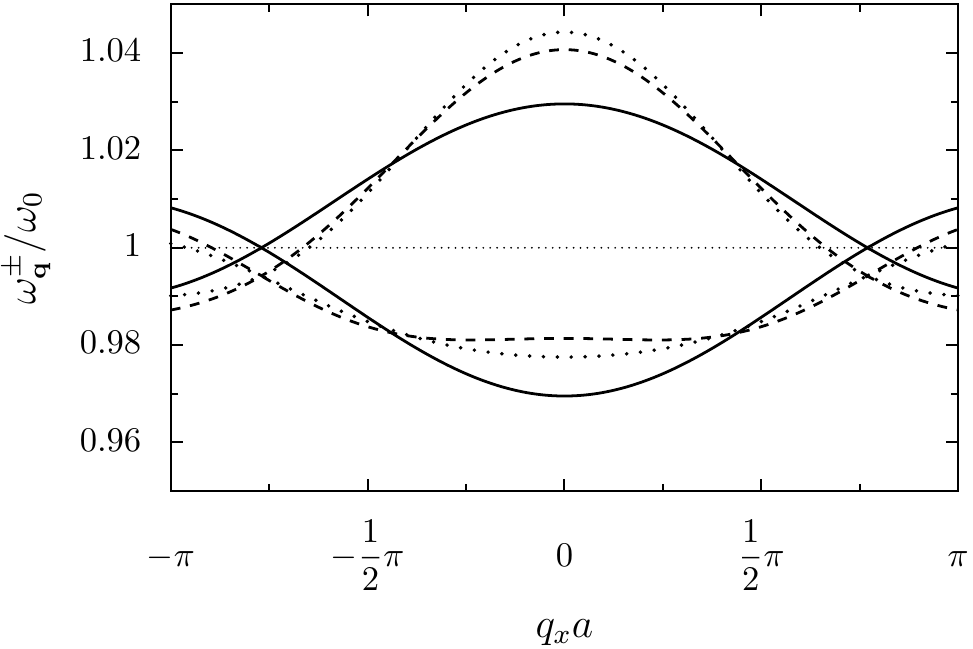}
\caption{\label{fig:warping}%
Collective plasmon dispersion for the out-of-plane mode ($\theta=0$) along the
$\mathrm{K}'\Gamma\mathrm{K}$ direction ($q_y=0$) 
including nearest (solid lines), next nearest (dashed lines) and third nearest
(dotted lines) neighbors. In the figure, $\Omega/\omega_0=0.01$.}
\end{figure}

In Fig.\ \ref{fig:warping}, we compare the collective plasmon dispersion for the out-of-plane 
mode ($\theta=0$) including only the nearest neighbors [solid lines in
Fig.\ \ref{fig:warping}, cf.\ Eq.\ \eqref{eq:dispersion_SM}] with the dispersion
obtained by taking into account also the next nearest [dashed lines, cf.\ Eq.\
\eqref{eq:dispersion_2}] and third
nearest neighbors [dotted lines, cf.\ Eq.\ \eqref{eq:dispersion_3}].
As can be seen from Fig.\ \ref{fig:warping}, the interactions beyond nearest
neighbors break the symmetry between the $+$ and $-$ branches, as it is the
case for graphene where the hopping between next nearest neighbors breaks the
particle-hole symmetry \cite{walla47_PR_SM, castr09_RMP_SM}. However, the dispersion close to the
$\mathrm{K}$ and $\mathrm{K}'$ points where the Dirac points stand is only slightly modified, merely introducing
a ``trigonal warping" of the Dirac cone \cite{castr09_RMP_SM}. Moreover, we estimated that 
the contribution of the nearest, next
nearest and third nearest neighbors to the total dipolar interaction energy is of the
order of \unit[75]{\%}.
We thus conclude that the physics close to the Dirac points described in the
main text should not be qualitatively modified by dipole-dipole interactions
beyond the ones involving nearest neighbors alone.

\section{Derivation of the condition for having a gapless plasmonic dispersion}
The dispersion \eqref{eq:dispersion_SM} is gapless if the modulus of the function 
\begin{equation}
\label{eq:f_q}
f_\mathbf{q}=\mathrm{e}^{\mathrm{i}q_ya/2}
\left(
\mathcal{C}_1\mathrm{e}^{-3\mathrm{i}q_ya/2}
+\mathcal{C}_2\mathrm{e}^{-\sqrt{3}\mathrm{i}q_xa/2}
+\mathcal{C}_3\mathrm{e}^{\sqrt{3}\mathrm{i}q_xa/2}
\right)
\end{equation}
vanishes, that is, if the function $f_\mathbf{q}=0$ itself for some value of
$\mathbf{q}$. This requires that
both real and imaginary parts of 
$\mathcal{C}_1\mathrm{e}^{-3\mathrm{i}q_ya/2}
+\mathcal{C}_2\mathrm{e}^{-\sqrt{3}\mathrm{i}q_xa/2}
+\mathcal{C}_3\mathrm{e}^{\sqrt{3}\mathrm{i}q_xa/2}
$ vanish, leading to the system of equations 
\begin{subequations}
\begin{align}
(\mathcal{C}_2+\mathcal{C}_3)\cos{\left(\frac{\sqrt{3}q_xa}{2}\right)}&=-\mathcal{C}_1\cos{\left(\frac{3q_ya}{2}\right)},
\\
\label{eq:qy}
(\mathcal{C}_2-\mathcal{C}_3)\sin{\left(\frac{\sqrt{3}q_xa}{2}\right)}&=-\mathcal{C}_1\sin{\left(\frac{3q_ya}{2}\right)}.
\end{align}
\end{subequations}
Adding the squares of these two equations, we arrive (if
$\mathcal{C}_2\mathcal{C}_3\neq0$) at
\begin{equation}
\label{eq:qx}
\sin^2{\left(\frac{\sqrt{3}q_xa}{2}\right)}=
\frac{(\mathcal{C}_2+\mathcal{C}_3)^2-\mathcal{C}_1^2}{4\mathcal{C}_2\mathcal{C}_3}.
\end{equation}
Such an equation only has a solution if
\begin{equation}
0\leqslant\frac{(\mathcal{C}_2+\mathcal{C}_3)^2-\mathcal{C}_1^2}{4\mathcal{C}_2\mathcal{C}_3}\leqslant1,
\end{equation}
which determines the polarizations of the electric field for which one has
a gapless plasmon dispersion.
To determine the locations in the first Brillouin zone of possible Dirac points,
one first solves for Eq.\ \eqref{eq:qx} to obtain $q_x$, and then obtain $q_y$ with Eq.\
\eqref{eq:qy} (if $\mathcal{C}_1\neq0$).


\end{document}